\documentclass{mem}
\usepackage{natbib}\usepackage{txfonts}\usepackage{balance}
\usepackage{graphicx}
\usepackage{subfigure}
\usepackage[a4paper]{hyperref}
\idline{75}{282}
\begin{document}
\def\teff{$T\rm_{eff }$}
\def\kms{$\mathrm {km s}^{-1}$}

%
%

\title{
Monitoring the radio spectra of selected blazars in the Fermi-GST era
}

\subtitle{The Effelsberg 100~m telescope covering the cm band}

\author{
E. \,Angelakis\inst{}, L. \,Fuhrmann\inst{}, N. \,Marchili\inst{}, T.\,P.~\,~Krichbaum\inst{}     
\and J.\,A.~\,~Zensus\inst{}}

  \offprints{E. Angelakis}

\institute{
Max-Planck-Institut f\"ur Radioastronomie, Auf dem H\"ugel 69, 53121 Bonn, Germany
\email{angelaki@mpifr-bonn.mpg.de}
}

\authorrunning{Angelakis et al.}

\titlerunning{Monitoring the radio spectra of selected blazars in the Fermi-GST era}

\abstract{The analysis of the spectral energy distribution variability
  at frequencies from radio to TeV is a powerful tool in the
  investigation of the dynamics, the physics and the structure
  evolution occurring in the most exotic flavour of active galaxies,
  the blazars. In particular, the presence of {\sl Fermi-GST} is
  providing a unique opportunity for such studies delivering
  $\gamma$-ray data of unprecedented quality. Here we introduce a
  monitoring program that has been running at the Effelsberg 100~m
  telescope since January 2007, underpinning a broad multi-frequency
  collaboration of facilities that cover the band from radio to
  infrared. Sixty one selected blazars are observed monthly between
  2.64 GHz and 43 GHz. The calibration accuracy is better than a few
  percent as it is demonstrated with some preliminary examples.
   }
\maketitle{}

\section{Introduction}
\begin{table}[h]
  \caption{The monitored sources.}     
  \label{angelakis_gamma_tab1}  
  \begin{center}                    
  \begin{tabular}{lll} 
    \hline \\                 
    \multicolumn{3}{c}{Target sources}\\        
    \hline                         
PKS\,0003$-$066  &TXS\,0814$+$425    &OE\,355  \\ 
PKS\,0215$+$015  &1E\,0317.0$+$18    &OJ\,248  \\
PKS\,0235$+$164  &H\,1426$+$428      &OJ\,287  \\ 
PKS\,0238$-$084  &1ES\,1959$+$650    &OP\,313  \\ 
PKS\,0336$-$019  &1ES\,2344$+$514    &3C\,84   \\             
PKS\,0420$-$014  &1ES\,1544$+$820    &3C\,111  \\           
PKS\,0528$+$134  &1ES\,0502$+$675    &3C\,120  \\           
PKS\,0735$+$178  &0059$+$581         &3C\,273  \\                   
PKS\,0748$+$126  &0219$+$428         &3C\,279  \\                   
PKS\,1038$+$064  &1128$+$592         &3C\,345  \\                   
PKS\,1127$-$145  &Mkn\,180           &3C\,371  \\           
PKS\,1406$-$076  &Mkn\,421           &3C\,446  \\           
PKS\,1510$-$089  &Mkn\,501           &3C\,454.3\\          
PKS\,1730$-$130  &4C\,21.35   &CTA\,102\\                  
PKS\,2155$-$152  &4C\,28.07   &Cyg\,A  \\                
PKS\,2155$-$304  &4C\,31.63   &BL\,LAC \\                
PKS\,2345$-$16   &4C\,38.41   &WCom    \\                
S5\,0836$+$71    &4C\,56.27   &M87     \\                
S5\,1803$+$78    &4C\,47.08   &OS\,319 \\                
S5\,0716$+$71    &TON\,0599   &\\          
S4\,0954$+$65    &NRAO\,150   &\\          
    \hline                                  
\end{tabular}           
\end{center}
\end{table}
Being among the most dramatic manifestations of the activity induced
in the nuclei of active galaxies, blazars comprise a unique probe of
the exotic physics at play in such systems. Their phenomenology is
dominated by extreme characteristics such as high degree of linear
polarisation, intense variability -- both in total power and
polarisation -- at all wavebands, apparent motions of features in
their radio structure, which are often highly superluminal and
brightness temperatures exceeding the Compton limit \cite[see
e.g.][]{angelakis_gamma_Urry1999APh}. This violent behaviour is
attributed to relativistic jets oriented very close ($\le
20^{\circ}$to $30^{\circ}$) to the line-of-sight \cite[see
e.g.][]{angelakis_gamma_Urry1995PASP}. Their spectral energy
distribution (SED) is characterised by the presence of two peaks. The
lower energy one, spreading from radio to far ultraviolet and soft
X-rays, is believed to be due to relativistic electrons emitting via
the synchrotron mechanism. The high energy part, reaching
up to TeV energies is believed to be produced by synchrotron
self-Compton or external Compton scattering.

Despite the overall understanding, many details remain unclear and are
subject to models attempting their clarification. For instance,
several ideas have been put forth to explain the origin of their
variability. Two of them are like the shock-in-jet model
\citep[e.g.][]{angelakis_gamma_Marscher1985ApJ,angelakis_gamma_Aller1985ApJ,angelakis_gamma_Marscher1996ASPC}
and the relativistic plasma shells
\citep[e.g.][]{angelakis_gamma_Spada2001MNRAS,angelakis_gamma_Guetta2004AnA}. Alternatively,
it has been suggested that light-house effect could be causing
variability in cases of rotating helical jets or the helical
trajectories of plasma elements
\citep[e.g.][]{angelakis_gamma_Begelman1980Natur,angelakis_gamma_Camenzind1992AnA}.

A powerful tool in the investigation of these details and for
understanding the dynamics, the physics and the structure of the
radiating regions is the analysis of the variability of their SEDs at
frequency ranges as broad as possible (from radio to TeV). Such an
approach can shed light on linear scales inaccessible even to
interferometric techniques and most importantly discriminate between
different variability scenarios. Particularly, the presence of {\sl
  Fermi-GST} is providing a unique opportunity for these studies
providing the $\gamma$-ray data of unprecedented quality.

To benefit fully from the newly flying {\sl Fermi-GST}
telescope, a broad collaboration of facilities that aim at
understanding the blazar phenomenon via the broad-band SED variability
monitoring has been initiated by
\cite{angelakis_gamma_Fuhrmann2007AIPC}. A sample of 61 selected
sources are monitored monthly and in tight coordination with other
facilities covering the cm, mm, optical and infrared bands (Fuhrmann
et al. in prep.). Here we present a very brief introduction of the
activities in the centimetric band with the 100~m Effelsberg telescope
and show some preliminary examples.

\section{The 100~m telescope monitoring program}
Observationally, the goal of the currently discussed project is the
monthly radio spectrum monitoring for a sample of the 61 sources shown
in table~\ref{angelakis_gamma_tab1} with the Effelsberg 100~m
telescope. The broad collaboration involves also mainly the 30~m IRAM
telescope and the 40~m Owens Valley telescope
\citep{angelakis_gamma_Fuhrmann2007AIPC}. The former is covering the
band between roughly 86 GHz and 270 GHz whereas the latter is focusing
on 15 GHz alone to monitor a few hundred sources per day and acquire
information about their variability index. The time difference between
the Effelsberg and IRAM observations is of order of one to maximum a
few days. The monitoring program at Effelsberg has been running
continuously since January 2007 providing a remarkable flow of data.

\section{The sample}
The sources in table~\ref{angelakis_gamma_tab1} have been selected
from the ``high-priority blazars'' list of the {\sl Fermi-GST} AGN
team. There is a remarkable overlap with other studies such as
2\,cm-VLBA survey/MOJAVE program
\citep{angelakis_gamma_Kellermann1998AJ,angelakis_gamma_Zensus2002AJ,angelakis_gamma_Kellermann2004ApJ},
the Boston 43 GHz VLBI survey \citep{angelakis_gamma_Jorstad2001ApJS},
the IRAM 30-m telescope polarisation monitoring program, the OVRO
monitoring program at 15 GHz and other multi-frequency campaigns.

\section{Observations and data reduction}
For our monitoring program the receivers at 2.64~GHz, 4.85~GHz,
8.35~GHz, 10.45~GHz, 14.60~GHz, 23.05~GHz, 32.00~GHz and 42.00 GHz
have been employed. Almost all of them (except for 32.00 GHz) deliver
polarisation information, which provides another useful tool in the
exploration of blazar physics.

The measurements are conducted with the newly installed adaptive
secondary reflector characterised by low surface RMS that induces
higher sensitivity (up to 50 \% increase at 42.00 GHz). The so-called
``cross-scan'' observing technique has been applied.
The advantage of this method is mainly the fact that it allows the
direct detection of cases of confusion as well as the correction for
pointing offset errors. The individual spectra are measured
quasi-simultaneously within $\le 40$ min to guarantee that they are
free of source variability of time-scales longer than that.

Considerable effort is put in applying some necessary
post-observational corrections to the raw data. Namely, (a) pointing
offset correction, (b) gain correction, (c) opacity correction and (d)
sensitivity correction
\cite[see][]{angelakis_gamma_Angelakis2008}. The sensitivity
correction is done with reference to standard calibrators
(e.g. table~\ref{angelakis_gamma_t2}). There, the average flux
densities are given along with the modulation index, $m$, over a
number of measurements ($m=100\cdot \frac{rms}{<S>}$). The modulation
index serves as a measure of the source variability. Hence, this table
is a demonstration of the system repeatability and the achievable
precision on the assumption that the calibrators are intrinsically
non-variable.
\begin{table}[]
  \caption{Average flux densities $S$ in Jy for two of our 
    main calibrators after preliminary analysis along with their modulation 
    index $m$ in percentage, as a function of $\nu$ in GHz. 
    $m$ serves as a pragmatic measure of the  
    system repeatability. For $\nu\ge 23.05$, NGC~7027 appears slightly
    extended for the Effelsberg beam. In column 4 we give its observed
    flux density $S'$ convolved with the observing beam. In column 6 we show
    the beam correction factor $\alpha$ in percentage, for a $10\times12''$ disk-like
    brightness distribution. In column 7 we list the total flux density
    $S = \left(1+\alpha/100\right)\cdot S'$.
  }
\label{angelakis_gamma_t2}
  \begin{center}
    \begin{tabular}{ccccccc}
      \hline
      \\
      $\nu$ &\multicolumn{2}{c}{3C\,286} &\multicolumn{4}{c}{NGC\,7027}\\ 
      &$S$ &$m$ &$S'$ &$m$ &$\alpha$ &$S$\\ 
      \hline
      \\
      2.64  &10.7 &0.7 &3.7 &1.1 & & \\  
      4.85  &7.5  &0.6 &5.5 &0.2 & & \\
      8.35  &5.2  &0.7 &5.9 &0.6 & & \\
      10.45 &4.5  &1.8 &5.9 &2.3 & & \\
      14.60 &3.5  &1.9 &5.8 &1.5 &   &    \\
      23.05 &2.5  &2.8 &5.3 &2.6 &6.6& 5.7 \\
      32.00 &1.9  &2.8 &5.2 &2.3 &5.6& 5.5\\
      42.00 &1.4  &2.1 &5.0 &5.0 &7.0& 5.4
      \\
      \hline

    \end{tabular}
  \end{center}
\end{table}

\section{Spectra}
\begin{figure}[t!]
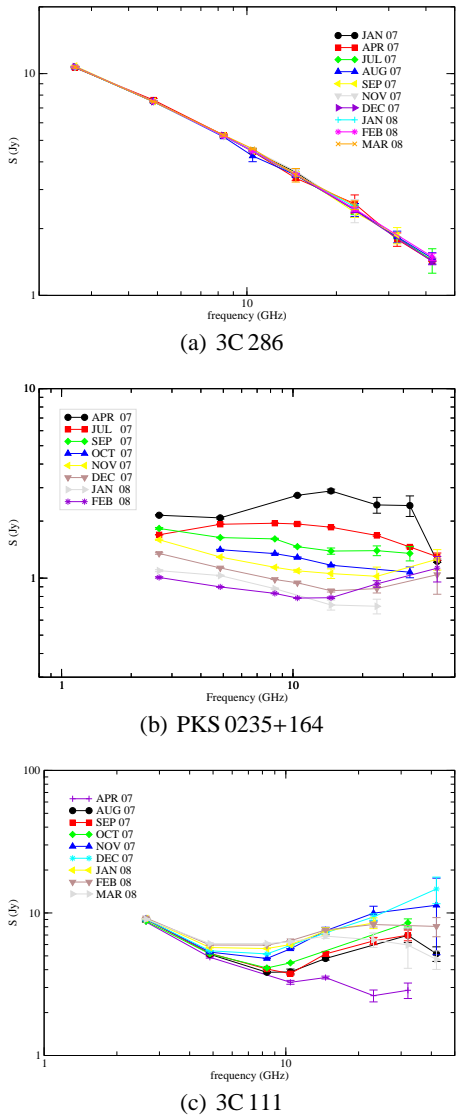

\subfigure[\footnotesize 3C\,286]
  {
    \resizebox{0.9\hsize}{!}{\includegraphics[clip=true]{angelakis_gamma_3C286.eps}}
    \label{angelakis_gamma_3C286}
  }
  \\
  \subfigure[\footnotesize PKS\,0235$+$164]
  {
    \resizebox{0.9\hsize}{!}{\includegraphics[clip=true]{angelakis_gamma_0059.eps}}
    \label{angelakis_gamma_0235}
  }
  \\
  \subfigure[\footnotesize 3C\,111]
  {
    \resizebox{0.9\hsize}{!}{\includegraphics[clip=true]{angelakis_gamma_3C111.eps}}
    \label{angelakis_gamma_3C111}
  }
  \caption{\footnotesize Examples of sources that exhibit intense
    spectrum variability in the cm band. The top panel shows the
    spectrum variability of one of the main calibrators namely 3C\,286
    for comparison. 3C\,286 is calibrated with calibration
    factor averaged over all used calibrators.}
\label{angelakis_gamma_f2}
\end{figure}

In figure~\ref{angelakis_gamma_f2} are collected two preliminary
examples of spectra measured with Effelsberg. The comparison of these
spectra with that of 3C\,286 in figure \ref{angelakis_gamma_3C286}
illustrates the significance of the observed variability. Remarkably, intense
variability appears already in the cm band. Its correlation with that
in other bands or possible changes in the structure of interferometric
imaging is under examination. It is important to mention that the
admittedly larger uncertainties at the higher frequencies is solely
due to atmospheric variations.

\section{Discussion}
Provided that the calibrators are presumably not intrinsically
variable, figure~\ref{angelakis_gamma_3C286} serves as an excellent means of
estimating the system repeatability. In table~\ref{angelakis_gamma_t2}
the calibration precision is of the order of a few percent. Factors
that induce such variations are primarily tropospheric variations,
opacity effects, confusion and system instabilities. At higher
frequencies (23.05 GHz and above), the atmospheric opacity becomes
particularly important.

The plots in figure~\ref{angelakis_gamma_f2} show how intensive and
fast the variability of the spectrum may be. This very fact urges for
dense frequency coverage especially of the millimetre regime from
where the evolution of a flaring event in the radio band is expected
to start. The evolution of the spectrum over time is among the
first-priority studies. That requires tight coordination of the
different participating stations as has been managed so far. The
activity seen in the centimetre band is noteworthy. Whether this
activity is correlated with that at other bands and especially that at
high energies, or with changes in the structure is one of the first
questions to be addressed. The behaviour of the polarisation when such
an event occurs is also among the most important issues to be
investigated.

\begin{acknowledgements}
  Based on observations with the 100~m telescope of the MPIfR
  (Max-Planck-Institut f\"ur Radioastronomie). 
\end{acknowledgements}

\bibliographystyle{aa}

\end{document}